# Structural characterization and bonding energy analysis for plasma-activated bonding of SiCN films: A reactive molecular dynamics study

Juheon Kim[a], Minki Jang[a], Junhyeok Park[a], Byungjo Kim[b,*], Hayoung Chung[a,*]

[a] *Department of Mechanical Engineering, Ulsan National Institute of Science and Technology, 50 Unist-gil, Ulsan 44919, South Korea*

[b] *Graduate School of Semiconductor Materials and Devices Engineering, Ulsan National Institute of Science and Technology, 50 Unist-gil, Ulsan 44919, South Korea*

**Abstract**

Plasma-activated bonding of SiCN films offers high bonding strength at the hybrid-bonding interface, thereby enhancing mechanical reliability. Although experimental studies have shown that the interfacial bonding properties of SiCN films vary with SiCN composition and plasma treatment parameters, a clear correlation between these parameters and the resulting bonding properties has not yet been established. This study presents an atomistic investigation of SiCN–SiCN plasma-activated bonding with controlled SiCN composition and plasma fluence, which performs $O_2$ plasma surface activation, surface hydroxylation, direct bonding, post-bonding annealing, and debonding using reactive molecular dynamics. The structural characterization of the plasma-activated SiCN surface, including density of various covalent bonds and surface roughness, exhibits composition- and plasma fluence-dependent chemical and morphological modification. Bonding energy evaluated from atomic traction–separation responses in cohesive zone volume elements (CZVE) during debonding simulations shows a positive correlation with the interfacial Si–O–Si density. Since the interfacial Si–O–Si density reflects the combined effects of these chemical and morphological modifications, the dependence of bonding energy on composition and plasma fluence is successfully elucidated by the structural characterization. These results establish an atomic-level material–process–property relationship and offer practical guidance for optimizing SiCN composition and plasma treatment parameters for SiCN-SiCN plasma-activated bonding.

**Keywords:** Electronic packaging, Plasma-activated bonding, Dielectric materials, Bonding energy, Reactive molecular dynamics, Cohesive zone model

## 1. Introduction

Driven by the increasing demand for semiconductor form-factor miniaturization and high-density integration, advanced packaging has gained prominence. In this context, hybrid bonding is a key technology for vertically stacking memory chips.[1–4] The dielectric–dielectric direct-bonded interface, which occupies most of the hybrid bonding area, serves as a critical component that sets the planarity and spacing required for subsequent metal–metal contact and that provides electrical isolation and mechanical load transfer. It also offers advantages over solder-based approaches that introduce foreign substances and limit fine-pitch scaling.[5,6]

As a conventional method for dielectric–dielectric bonding, $SiO_2$–$SiO_2$ direct bonding is typically used.[7–9] Despite its maturity, $SiO_2$–$SiO_2$ bonding can be constrained by void formation at the bonding interface during post-bonding annealing, which lowers the attainable bonding strength.[10–12] Recently, amorphous SiCN (a-SiCN) has emerged as an alternative dielectric for direct bonding. Relative to $SiO_2$ and other dielectrics, SiCN–SiCN direct bonding offers favorable thermo-mechanical stability, the ability to attain high interfacial properties at reduced temperatures after plasma treatment and surface hydroxylation, and demonstrates effectiveness as a copper diffusion barrier at the Cu-dielectric contact region due to the Cu pad misalignment.[13] Under comparable plasma activation, a-SiCN can also exhibit a higher areal density of dangling bonds than $SiO_2$, providing more precursors for silanol (Si–OH) group formation and subsequent Si–O–Si linkages.[14] Several properties of a-SiCN vary with its composition, and the constituent elements play different roles at dielectric bonding interfaces: under plasma surface activation, silicon and carbon elements provide chemical bonding sites by forming dangling bonds that directly participate in interfacial bonding, whereas nitrogen acts as a diffusion barrier suppressing Cu migration into the dielectric layer at Cu–dielectric interface.[15,16]

Experimental studies on SiCN–SiCN direct bonding report successful bonding after $O_2$ or $N_2$ plasma treatment, surface hydroxylation, room-temperature bonding, and post-bonding annealing, with interfacial properties that depend on composition and plasma fluence.[17–19] Reported structural characterizations include plasma-induced changes in surface chemistry, generation and hydroxylation of dangling bonds, formation of Si–O–Si during contact and annealing, and morphological changes such as surface roughness growth and local inhomogeneous topography, together with comparisons of bonding properties across process conditions. Although differences in bonding energy with a-SiCN composition have been observed experimentally, a clear correlation of bonding energy as a function of stoichiometry has not been firmly established. The interplay among plasma treatment conditions, hydroxylation conditions, annealing method, and the resulting chemical and morphological surface characteristics makes it difficult to draw a general material-process-property correlation for SiCN–SiCN plasma-activated bonding.

To clarify the material-process-property correlation, atomistic-level phenomena, such as plasma-surface interactions and interfacial bonding, need to be investigated with controlled modeling and process parameters, and with quantitative structural characterization that captures both chemical modification and morphological modification. These requirements motivate the use of reactive molecular dynamics (MD) simulations, in which a reactive force field (ReaxFF) allows bond formation and dissociation based on bond order calculation to occur during the atomistic simulation without explicit bond definitions or predefined reaction pathways.[20,21] Prior MD studies have examined plasma surface activation and subsequent surface hydroxylation of a-SiCN, reporting composition- and fluence-dependent changes in the bonding network and dangling-bond distributions, as well as surface-morphology descriptors.[22,23] Kim et al.[23] performed structural characterization of an atomistic model for a-SiCN films after plasma-induced activation and DI-water rinsing, but the SiCN–SiCN direct bonding and bonding energy evaluation were presented as future work. However, to the best of authors' knowledge, direct evaluation of SiCN–SiCN bonding properties, such as bonding energy and bonding strength at the bonded interface, has not yet been carried out through MD simulations.

In this study, we employed reactive molecular dynamics simulation to perform a SiCN-SiCN plasma-activated bonding sequence, including the $O_2$ plasma treatment, surface hydroxylation, direct bonding, and post-bonding annealing. The structural characterization of the SiCN surface and bonding interface was quantified throughout the simulation processes in terms of chemical and morphological modifications, including bond-distribution changes, areal densities of dangling bonds, Si–OH and Si–O–Si, and surface roughness. Subsequently, debonding simulations of the bonded SiCN films were conducted to obtain atomic traction–separation (T–S) responses in cohesive zone volume elements (CZVE), and the T–S law was represented by an exponential cohesive law to evaluate bonding energy. Finally, the resulting bonding energy was elucidated by the structural characterization to identify the correlation between composition- and plasma fluence-dependent surface modification and bonding energy.

## 2. Computational methods

### 2.1. Amorphous SiCN model description

In this study, we first constructed a-SiCN single layer models having different compositions, which undergo the surface treatments, such as plasma surface activation and hydroxylation. The models were then used to build the bilayer models.

To investigate bonding behavior and properties based on SiCN stoichiometry, three models with different compositions were constructed: SiCN, $SiC_2N$, and $SiC_3N$. Every model consists of 27,000 atoms, and the number of atoms of each element is proportional to the composition ratio. A unit cell size is 8 nm x 8 nm x 5 nm, and the atoms were randomly placed inside the periodic unit cell. Only the carbon ratio varies among the models to reflect the stoichiometric inhomogeneity that develops in a-SiCN as carbon-rich domains form with increasing carbon content.[24]

A two-step interatomic potential approach was employed to construct the a-SiCN single layer model, which involves two different modeling procedures sequentially under each interatomic potential: (1) Tersoff potential[25,26] and (2) reactive force field (ReaxFF)[27]. In this approach, the Tersoff potential was first used to generate a structurally realistic amorphous SiCN network that reproduces the characteristic features of the SiN matrix and carbon-rich (C-rich) nanodomains. Subsequently, the ReaxFF potential was applied to simulate bond formation and dissociation during the following thermo-mechanical processes. This sequential procedure allows the Tersoff potential to generate a structurally realistic amorphous SiCN network, followed by the ReaxFF potential that captures chemical reactions during process simulations.[22]

A melt-quench process using the Tersoff potential, which induces amorphization of the material by rapidly melting and cooling to freeze the disordered atomic configuration into a stable amorphous phase, as reported in a previous study.[24] The model was equilibrated at 0.1 K for 20 ps and heated to 8000 K for 20 ps. Then, the structure was quenched to 3000 K for 40 ps and relaxed for 200 ps. The structure was cooled from 3000 K to 300 K for 1 ns, and a bulk a-SiCN model was finally obtained after relaxation at 300 K for 10 ps. A timestep of 0.5 fs was selected for the simulation process.

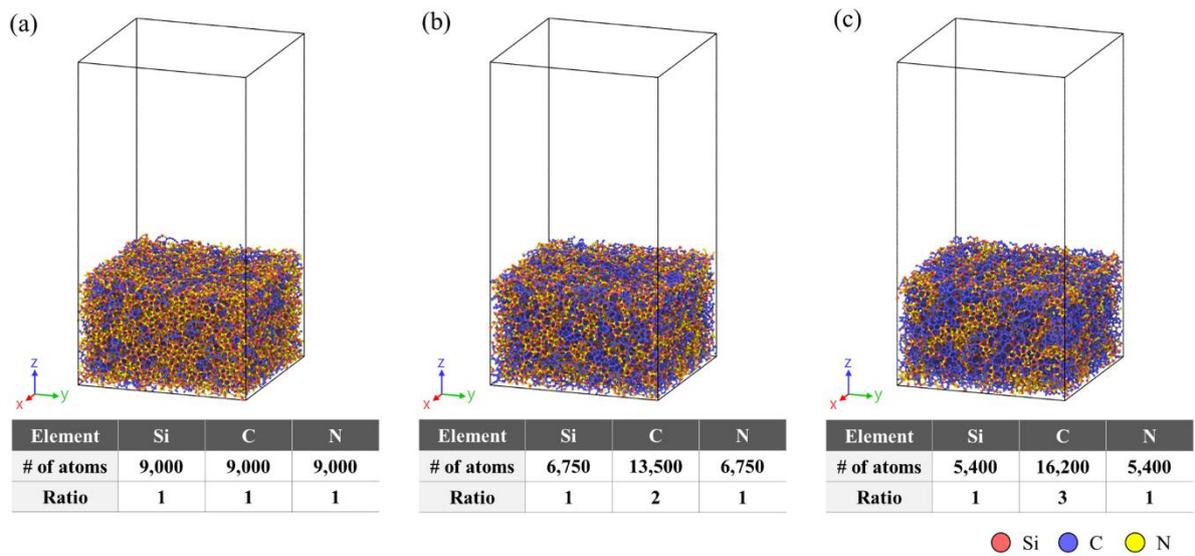

Fig. 1. Atomistic models of a-SiCN single layers with different compositions: (a) SiCN, (b) SiC$_2$N, and (c) SiC$_3$N. Each structure was equilibrated using a two-step interatomic potential approach (Tersoff-ReaxFF). The number of atoms and composition ratio for each model are summarized in the tables below.

Subsequently, ReaxFF was used as an interatomic potential, and the bulk a-SiCN model was transformed into a slab with a finite thickness of 5 nm. The bulk a-SiCN model was relaxed through the NVT ensemble simulation for 40 ps at 300 K, and the NPT ensemble simulation was conducted for 20 ps at 300 K and 0 MPa. Then, the upper boundary of the simulation box along the z-direction was moved up 10 nm to form the top surface. Non-periodic boundary conditions were applied in the z-direction, and several layers within 4 Å-thickness from the bottom were fixed, while periodic boundary conditions were still applied in x- and y-direction. Finally, the equilibrated a-SiCN single layer model was obtained after the relaxation process through the NVT and NPT ensemble simulations in series for 20 ps at 300 K and 0 MPa, as shown in **Fig. 1**. A timestep of 0.25 fs was used throughout the simulations using ReaxFF.

### 2.2. Simulation of the plasma-activated bonding process

The plasma-activated bonding process of a-SiCN films was simulated in the following sequence: plasma surface activation, surface hydroxylation, direct bonding, and post-bonding annealing, as illustrated in **Fig. 2**. The surface of the a-SiCN single layer was first subjected to $O_2$ plasma treatment to induce plasma surface activation, followed by surface hydroxylation through interaction with DI water to form Si–OH groups. Two hydroxylated surfaces were then brought into contact to achieve direct bonding, and a subsequent post-bonding annealing process was performed to promote interfacial Si–O–Si formation.

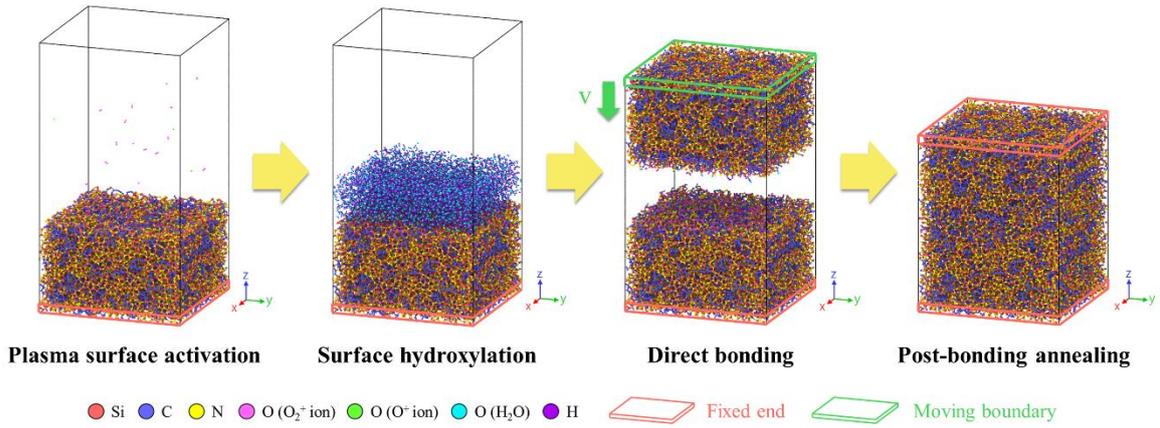

Fig. 2. Simulation procedure of the plasma-activated bonding process for a-SiCN films, including plasma surface activation, surface hydroxylation, direct bonding, and post-bonding annealing.

### 2.2.1 Plasma surface activation of a single SiCN layer

To induce direct bonding of dielectric films, the surfaces should be activated energetically by the plasma surface activation. This process causes the dissociation of existing chemical bonds or the formation of new chemical bonds by ion bombardment to the surface, resulting in plasma-induced surface modification.

In this work, $O_2$ plasma treatment was performed to activate the surface of the a-SiCN single layer model, as shown in **Fig. 1**. The $O_2^+$ and $O^+$ ions that are $O_2$ plasma species were inserted at randomly selected positions inside the plasma depositing region 70–80 Å above the a-SiCN surface. To avoid the overlapping of newly generated oxygen ions across the periodic boundaries in x- and y-direction during the plasma treatment simulations, the deposition region in the xy plane was constrained to be 5 Å away from each lateral boundary. We assigned constant energy to ion species ($O_2^+$ and $O^+$ ions), following Hahn et al.[22] Three plasma fluence conditions, which indicate the number of ions passing through the unit area of the surface per unit time, were investigated, and their ion energy, flux, and fluence are tabulated in **Table 1**. To reflect the difference in the fluxes of $O_2^+$ and $O^+$ ions, each ion was inserted into the simulation box every 25 fs and 105 fs, respectively. These ion insertion intervals were intended to reflect the continuous effects of the plasma reaction while preventing ion–ion interactions before reaching the substrate.[28] The inserted ions were vertically toward the a-SiCN surface with predefined ion energy. During plasma treatment, the a-SiCN substrate was relaxed by the NVT ensemble simulation at 300 K with a Berendsen thermostat, which prevents the undesired temperature increase of the substrate due to the kinetic energy of impinging ions. After the plasma treatment, the activated a-SiCN film was relaxed at 300 K by using the NVT ensemble simulation for 100 ps. Although the plasma treatment and relaxation process were complete, several residues desorbed from the substrate, or non-bonded ion species remained in the vacuum slab above the substrate. As there was no further interaction between the residues and substrate, the residues were removed from the simulation box, and an additional relaxation process was performed for 20 ps.

**Table 1.** Process parameters for plasma surface activation simulation

| Ion species | Ion energy [eV] | Plasma flux [Ion #/nm²·ps] | Plasma fluence [Ion #/nm²] | | |
|---|---|---|---|---|---|
| | | | Low | Medium | High |
| $O_2^+$ | 75.2 [22] | 0.6514 | 8.143 | 10.178 | 12.214 |
| $O^+$ | 75.6 [22] | 0.1551 | 1.939 | 2.423 | 2.908 |

### *2.2.2 Surface hydroxylation of a single SiCN layer*

The plasma-treated a-SiCN surface undergoes hydroxylation through rinsing with deionized (DI) water to generate hydrophilic functional groups, which play a crucial role in the direct bonding interface, as these groups form interfacial bonds, such as siloxane bonds (–Si–O–Si–), through the condensation reaction between a-SiCN films.

A model consisting of DI water molecules was prepared separately, with a 2 nm thickness and the same lateral dimension as the a-SiCN film. The model includes 4,155 $H_2O$ molecules to ensure 1 g/cm³ density. Finally, the model was equilibrated by using the NVT ensemble simulation for 20 ps at 300 K. The equilibrated DI water model was placed 4 Å above the plasma-activated surface of a-SiCN substrate model.[29] The a-SiCN model exposed to DI water was relaxed at 300 K by using the NVT ensemble simulation for 50 ps, enabling surface hydroxylation through water-surface interaction. After the hydroxylation, residual water molecules that are not bonded to the a-SiCN substrate were eliminated from the simulation box.

### 2.2.3 Direct bonding of SiCN bilayer

A pair of a-SiCN thin films was placed facing each other, with modified surfaces through the plasma treatment and hydroxylation to construct a bilayer model of a-SiCN. The two a-SiCN thin films in the bilayer model were placed at a distance of 25 Å between their surface, which ensures an initial model without any interaction between the layers before the direct bonding process. The bilayer model was relaxed at 300 K by using the NVT ensemble simulation for 20 ps. After the relaxation process, direct bonding was achieved by displacing the upper layer-end until the initial contact between the upper and bottom surfaces. Then, the bilayer structure was relaxed at 300 K for 100 ps, which represents room-temperature bonding. Subsequently, a post-bonding annealing process was performed at 523 K to promote the formation of interfacial covalent bonds, which contribute to improved bonding strength.[30,31] The model relaxed at 300 K was heated up to 523 K with a 1 K/ps of heating rate, and remained at the annealing temperature for 200 ps. After the post-bonding annealing, the model was cooled down from the annealing temperature to room temperature with a 1 K/ps cooling rate, and it was relaxed at 300 K for 200 ps. Finally, with the fixed upper layer-end defined as the free-end along the z-axis only, the model was relaxed for 300 ps to release the compressive condition.

### 2.3 Bonding energy measurement

The debonding simulation was performed using ReaxFF, and the T–S response was analyzed within the cohesive zone volume element (CZVE). The bonding energy was then determined by fitting the atomic T–S data with an exponential cohesive law.

### 2.3.1 Debonding of the bonding interface

In this study, the mode I fracture behavior at the SiCN–SiCN bonding interface was analyzed using molecular dynamics simulation to evaluate the interfacial bonding energy, which corresponds to the mode I fracture energy typically measured in experiments using double cantilever beam (DCB) tests.[32,33] The debonding simulation was conducted using ReaxFF, which allows atomistic modeling of bond breaking during debonding at chemically bonded SiCN–SiCN interfaces.

The simulation was performed using SiCN–SiCN direct bonding models developed in **Section 2.2.3**. The structures were equilibrated at 300 K after undergoing direct bonding and post-bonding annealing processes. A quasi-static tensile strain was gradually applied in the z-direction to the top 4 Å region of the model at a strain rate of $10^9$/s, and tensile strain was applied every 400 timesteps for 200 ps, while the bottom 4 Å region was fixed. Periodic boundary conditions were imposed in the x- and y-directions, and interfacial debonding was induced through uniaxial tension along the z-direction.

### *2.3.2 Cohesive zone model of direct bonding interface*

The debonding behavior at the SiCN–SiCN interface exhibits nonlinear ductile characteristics arising from covalent bonding chains such as Si–O–Si, which resist interfacial separation. Such behavior can be modeled using cohesive zone model (CZM). In classical CZM theory, the interface is modeled as a zero-thickness cohesive layer. In MD simulations, however, a finite cohesive zone volume element (CZVE) must be defined to evaluate traction and separation. The CZVE thickness critically affects the calculated stress: if it is too thin, local fluctuations dominate due to insufficient atom count; if too thick, spatial averaging smears out localized fracture behavior. Hence, selecting an appropriate CZVE thickness is essential for accurately capturing the stress response within the fracture process zone. In this study, the CZVE was divided into upper and lower cells across the bonding interface, and its thickness along the z-direction was determined from the oxygen distribution depth induced by $O_2$ plasma treatment, corresponding to the fracture process zone where debonding initiates.

Due to the amorphous nature of a-SiCN, the interfacial region exhibits locally varying stoichiometries and bonding characteristics, primarily influenced by C-rich nanodomains that cause local stress variations. To capture this inhomogeneity, the interfacial plane was subdivided into multiple cells on the xy plane, as shown in **Fig. 3**, and the debonding behavior was analyzed for each cell individually. A 10 Å-wide edge region containing several voids was formed because the oxygen ion deposition region during plasma treatment was narrowed by 5 Å from the boundaries in the x- and y-directions, resulting in insufficient surface modification to establish stable interfacial bonding in that region. Although the edge-void was not intentionally designed, it effectively acted as a pre-crack, guiding the fracture process zone to coincide with the interface between the upper and lower cells. However, because interfacial bonding in the edge region was comparatively not well developed and caused significantly low traction during debonding, the edge region was excluded from the CZVE.

The atomic stress within each CZVE was calculated from the virial expression and the stress component $\sigma_{zz}$ was calculated as normal traction.[34,35] The separation displacement was calculated as the difference in average z-positions between atoms in the upper and lower cells of each CZVE, given by:

$$\delta = \delta(t) - \delta(0) = (\overline{z_u}(t) - \overline{z_l}(t)) - (\overline{z_u}(0) - \overline{z_l}(0)) \tag{1}$$

where $\delta(t)$ is the distance between the upper and lower cells at time $t$, $\overline{z_u}$ and $\overline{z_l}$ are the average z-position of the atoms in the upper and lower cells, respectively.

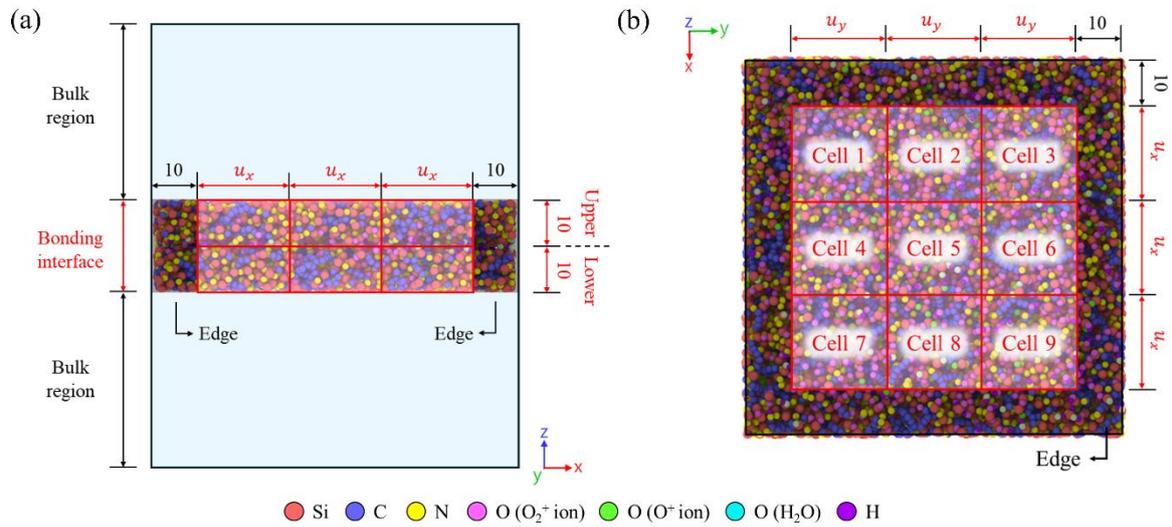

Fig. 3. Atomic configurations of the CZVE divided into an array of cells at the SiCN–SiCN interface. For the example of $3 \times 3$ cells: (a) front view showing upper and lower CZVE regions along the z-direction, and (b) top view illustrating the $3 \times 3$ array of cells on the interfacial plane. A unit of dimension is Å.

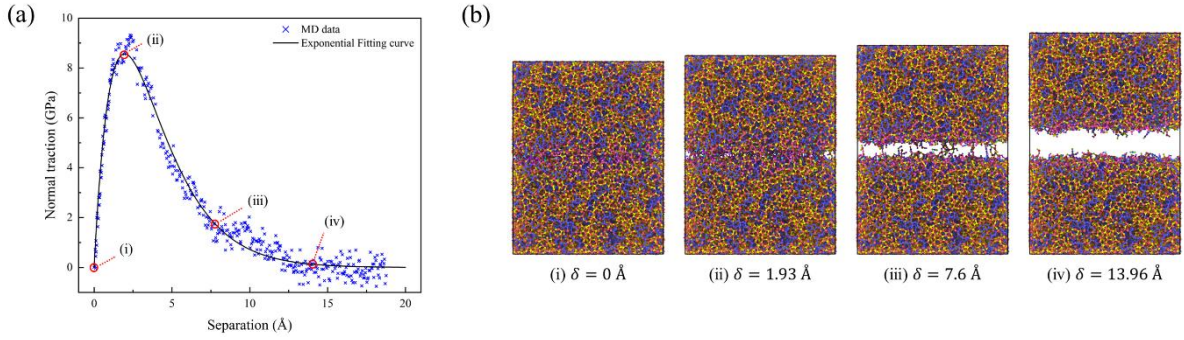

Fig. 4. (a) Traction-separation data obtained from the debonding simulation of bonding interface for a-SiCN with low plasma fluence and exponential fitting curve, and (b) corresponding atomic configurations of ( i )-(iv).

In this study, an exponential T–S law, as proposed by Needleman,[36] was used to fit the MD data. The T–S data obtained from molecular dynamics simulations were inherently discrete due to atomistic resolution, and therefore, the MD data were fitted to a continuous function to calculate the bonding energy, as shown in **Fig. 4**. This cohesive model is widely used to describe ductile interfacial fracture behavior and is expressed for Mode I loading conditions as follows:

$$T(\delta) = T_0 \frac{16e^2}{9} \frac{\delta}{\delta_C} \exp\left(-\frac{16e}{9} \frac{\delta}{\delta_C}\right) \qquad (2)$$

where $T_0$ is the maximum cohesive traction, and $\delta_C$ is the critical separation distance. In the least square method, the exponential function was parameterized using two fitting parameters: $T_0$ and $\delta_C$. These parameters were optimized by minimizing the root mean square error (RMSE) between the fitted curve and the T–S data obtained from molecular dynamics simulations. The area under the T–S curve represents the fracture energy, as follows:

$$\Gamma_c = \int_0^{\delta_f} T(\delta) d\delta \qquad (3)$$

where $\delta_f$ is the upper bound of the separation range for the exponential function. For the curve fitting and energy evaluation, the upper bound of the separation range was set to $\delta_f = 20\,\text{Å}$, since every examined debonding simulation exhibited complete interfacial separation with the traction decaying to zero below 20 Å. $\Gamma_c$ can be directly compared with the bonding energy obtained experimentally by DCB tests.[13,14,23,37] The resulting bonding energy was associated with Si–O–Si bond density, which was detected oxygen atoms simultaneously bonded to two silicon atoms within a ± 3.2 Å region about the interfacial mid-plane, corresponding to the maximum span of an Si–O–Si unit considering the Si–O bond length of 1.6 Å.[38]

## 3. Results and discussion

### 3.1 a-SiCN single layer models before surface modification

Constructing atomistic models of a-SiCN that reproduce the locally inhomogeneous composition is essential for investigating the relationship between local stoichiometry and bonding performance. Therefore, the atomistic models of single a-SiCN films were validated based on the radial distribution function (RDF) analysis, density evaluation, and identification of C-rich nanodomains to ensure that their physical properties are consistent with those of realistic amorphous structures.

**Fig. 5a–c** shows the RDF analysis for Si–C, Si–N, and C–C pairs in a-SiCN models with different compositions. The first peaks in the RDF results for Si–C, Si–N, and C–C pairs are observed at 1.85 Å, 1.87 Å, and 1.48 Å, respectively, which indicate the bond lengths of each pair. Such values remain constant regardless of composition. The bond lengths of Si–C and C–C are found to be in good agreement with theoretical and experimental results.[24,39,40] However, the Si–N bond length shows a deviation of about 0.1 Å from the experimental value. This deviation could arise because, unlike the Tersoff potential, where the attractive interactions of C–N and N–N were turned off to construct a physically realistic a-SiCN model,[41,42] the ReaxFF potential was used without such modification. Nevertheless, the bonding characteristics of N–N and C–N pairs, including the absence of $N_2$ molecule formation and few C–N bonds, are preserved, indicating that the deviation in Si–N bond length does not significantly impact the overall bonding network of a-SiCN. These RDF results verify that the atomic arrangement of the constructed a-SiCN models accurately represents the characteristic short-range order of amorphous SiCN.

**Fig. 5d** presents the density values with respect to carbon contents, and an inverse correlation between carbon content and density is observed. Such values and correlation agree well with the experimental results.[43] Since the density is closely related to material properties such as Young's modulus, hardness, thermal conductivity, and dielectric constant,[44] ensuring that the atomistic model yields a reasonable density compared to experimental values is essential for model validation. These results confirm that the constructed a-SiCN models reproduce the experimentally consistent mass density and compositional dependence, ensuring that the subsequent surface treatment and interfacial bonding simulations accurately reflect composition-dependent effects.

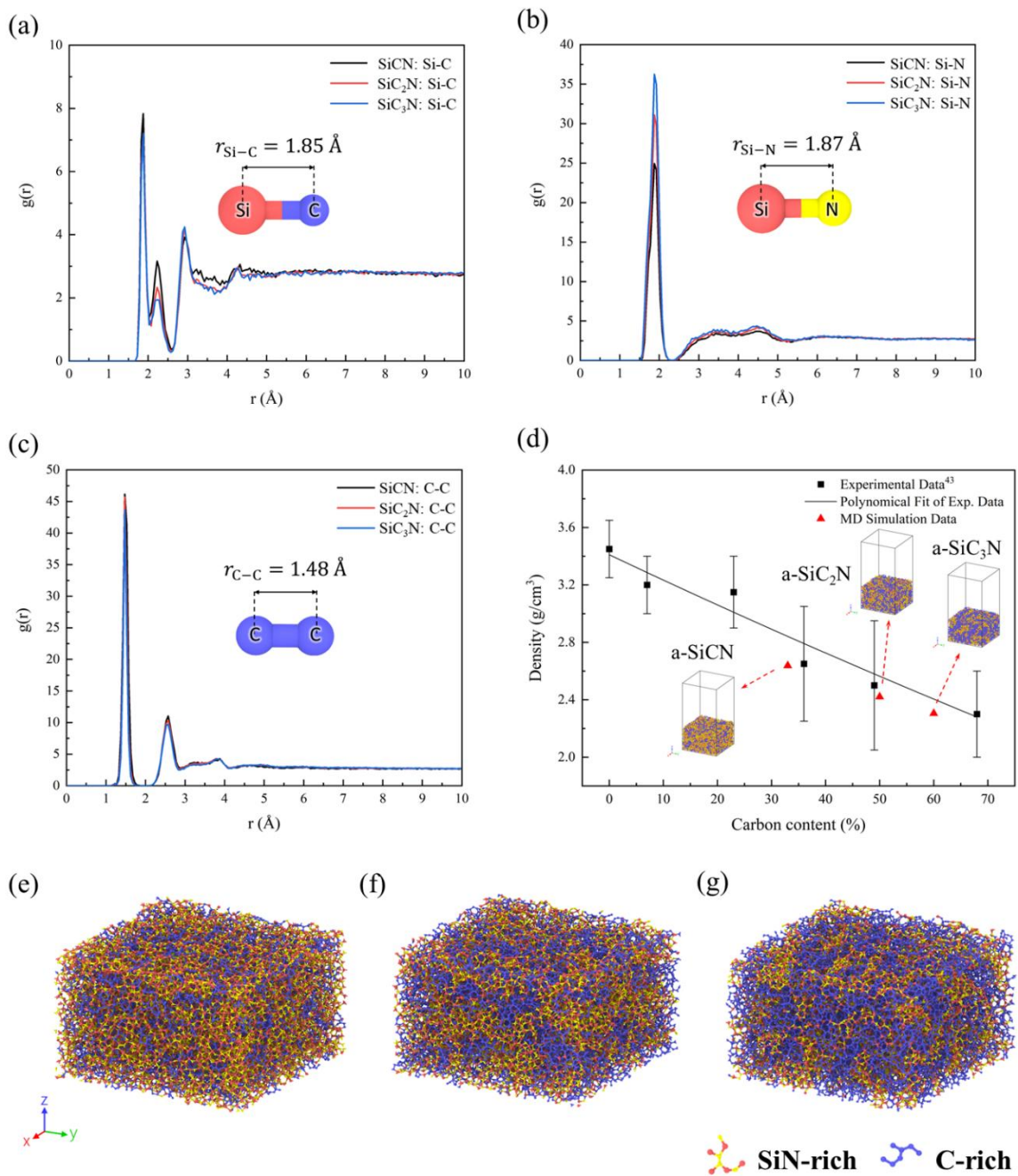

Fig. 5. (a–c) Radial distribution function analysis of Si–C, Si–N, and C–C for a-SiCN films. (d) Density of a-SiCN films with varying carbon content. Experimental values were measured by Lehmann et al.[43] (e–g) Front view of atomistic models of a-SiCN films with different compositions: SiCN, SiC$_2$N, and SiC$_3$N, illustrating SiN matrices and C-rich nanodomains.

As shown in **Fig. 5e–g**, the a-SiCN films exhibit distinctive microstructures composed of multiphase components, including SiN matrix and C-rich nanodomains, and the C-rich nanodomains grow in size as the carbon content increases. Such multiphase structures have been experimentally observed in a-SiCN films deposited by chemical vapor deposition (CVD)[45–47] as well as in polymer-derived a-SiCN.[48,49] The presence of C-rich nanodomains confirms the existence of stoichiometric inhomogeneity and suggests that the stoichiometry is not uniform throughout the structure. Since the local stoichiometric inhomogeneity is also reflected at the surface, it affects the surface modification during the surface treatment process. These results confirm that the constructed a-SiCN models maintain physically reasonable atomic configurations, providing a reliable structural basis for subsequent process simulations.

### *3.2 Structural characterization of surface-treated a-SiCN films*

### *3.2.1 Plasma-activated single a-SiCN layer*

In **Fig. 6a**, the plasma-treated SiCN film with 1:1:1 composition is shown as a representative atomistic model to visualize the vertical distribution of oxygen atoms. The deposited oxygen atoms are mostly located near the surface ($z \approx 50$ Å), while they are not observed beyond a certain depth, approximately $z \approx 35$ Å. **Fig. 6b** shows the depth-dependent atomic percent of oxygen (at% O) profiles under low plasma fluence for three different compositions. In all cases, at% O decreases monotonically with increasing depth and shows negligible dependence on composition at a given depth. **Fig. 6c** compares the at% O profiles in a-SiCN films (Si:C:N = 1:1:1) treated under different plasma fluence conditions. A consistent decrease in at% O with increasing depth is observed across all plasma fluence levels. It was found that higher plasma fluence results in higher at% O at each depth, which is attributed to the increased number of oxygen ions introduced at higher fluence, leading to more frequent collisions with the a-SiCN surface and a higher oxygen concentration within the film. For all SiCN compositions and plasma fluence conditions, the penetration depth of oxygen atoms is commonly limited to regions above $z = 35$ Å. This consistent penetration limit closely depends on the ion energy which is a process parameter of the plasma treatment.[50] In this study, a fixed ion energy was used during plasma treatment simulations, explaining the uniform maximum depth of oxygen penetration regardless of composition or plasma fluence. However, the amount of oxygen atoms in the region between $35\,\text{Å} \leq z < 40\,\text{Å}$ is negligible, as shown in **Fig. 6b** and **c**. Therefore, the effective depth of chemical modification due to plasma treatment extends up to 1 nm from the surface ($z \geq 40\,\text{Å}$).

**Fig. 7a** shows the bond order distribution of O atoms as a function of depth in a-SiCN films with varying compositions under low plasma fluence conditions. As the O atoms are located closer to the surface, the average bond order decreases, and the distribution broadens toward lower values. This reduction is particularly pronounced at the topmost surface, indicating the frequent presence of undercoordinated O atoms ($BO_{oxy} \approx 1$) due to the lack of neighboring atoms at the surface. In contrast, O atoms penetrated below the surface tend to form saturated bonds with surrounding atoms, which leads to higher bond orders, $BO_{oxy} \approx 2$. The difference in bond formation of O atoms with depth was also confirmed by comparing the atomic configurations and individual bond orders shown in **Fig. 7c** and **d**. Despite differences in the SiCN composition of the films, the depth-dependent bond order distribution is consistently observed, and the values show negligible variation.

**Fig. 7b** presents the bond order distribution of O atoms in a-SiCN films with a fixed composition (Si:C:N = 1:1:1) under varying plasma fluence conditions. Consistent with the observations in **Fig. 7a**, the surface region ($z \geq 50\,\text{Å}$) exhibits lower bond order values than the underlying bulk region ($35\,\text{Å} \leq z < 50\,\text{Å}$). With increasing plasma fluence, the mean value of bond order at the surface exhibits an increase, as indicated by the upward shift of the black dots in the surface box plots. The narrowing of the interquartile range (IQR, the range of 25–75%) under higher fluence conditions suggests that surface O atoms were more uniformly bonded, with a greater proportion of fully coordinated species. This is attributed to the increased number of incident ions as plasma fluence increases, which leads to higher atomic density near the surface and provides more neighboring atoms that facilitate saturated bond formation. Despite the increase in average bond order under higher fluence conditions, O atoms with low bond order values ($BO_{oxy} \approx 1$) persist across all fluence conditions. These values appear as statistical outliers, as shown in **Fig. 7b**, but physically correspond to undercoordinated atoms that represent unsaturated bonding states. Therefore, although higher plasma fluence promotes greater bond saturation of surface oxygen atoms, it also induces the formation of undercoordinated atoms that retain dangling bonds. These undercoordinated oxygen atoms act as chemically reactive sites on the plasma-treated surface, indicating that plasma treatment can generate unsaturated, reactive dangling bonds by modifying the surface bond network through ion bombardment.

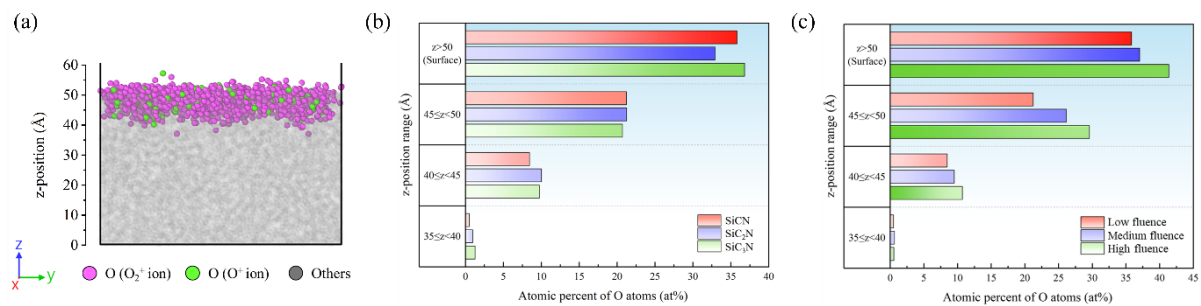

Fig. 6. Vertical distribution of oxygen atoms in $O_2$ plasma-treated a-SiCN films. (a) Side view of the a-SiCN model with 1:1:1 composition. Depth-dependent profiles of atomic percent of oxygen (b) under low plasma fluence for different a-SiCN compositions and (c) under different plasma fluence conditions for a single a-SiCN composition (Si:C:N=1:1:1).

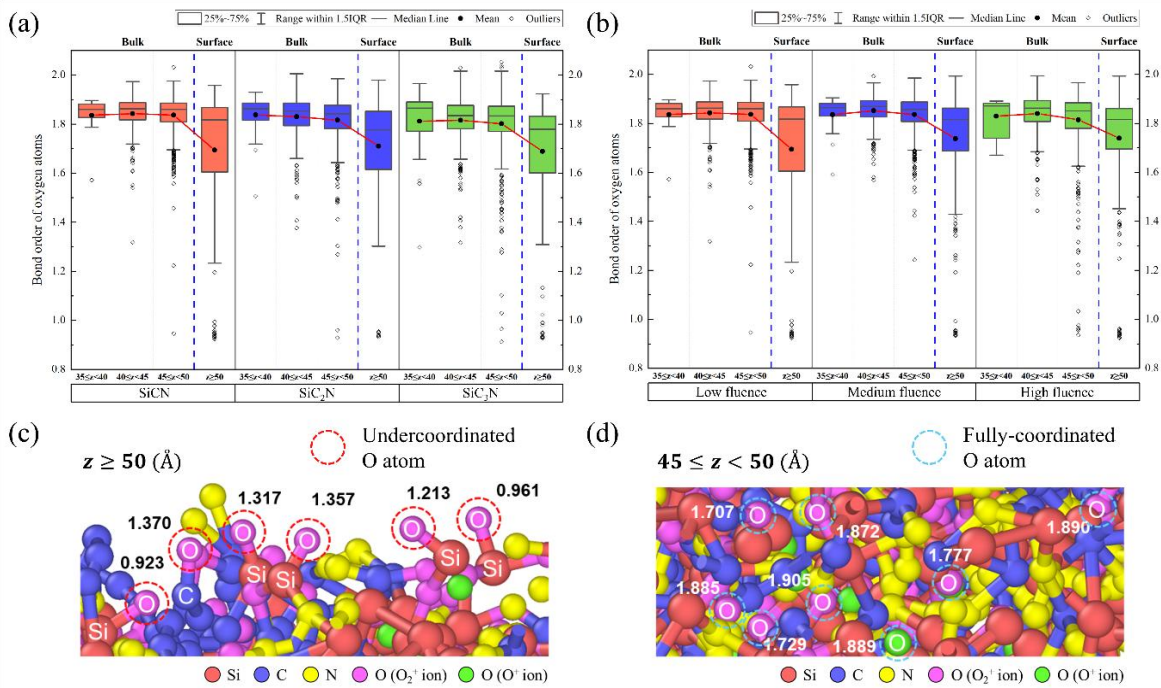

Fig. 7. Depth-dependent bond order distribution of oxygen atoms in $O_2$ plasma-treated a-SiCN films. Box plot of oxygen bond orders (a) under low plasma fluence for different a-SiCN compositions and (b) under varying plasma fluence conditions for a fixed composition (Si:C:N = 1:1:1). Atomistic configuration of the plasma-treated a-SiCN film with 1:1:1 composition under low plasma fluence, visualized for (c) the surface region (z ≥ 50 Å), and (d) the near-surface region (45 Å ≤ z < 50 Å).

In **Fig. 8a**, the formation of Si–O and C–O bonds is proportional to the Si and C content of the a-SiCN models, respectively. A higher Si content leads to an increased formation of Si–O bonds, which is accompanied by a greater reduction in the number of Si–Si and Si–N bonds, as these are the original bonding configurations of Si atoms. Similarly, the dissociation of C–C bonds is directly associated with the formation of C–O bonds. The reduction of Si–C bonds shows a composition-dependent change that parallels the formation of C–O bonds, which is attributed to the C-rich nanodomains, since Si–C bonds are mainly located at the interface between the SiN matrix and the C-rich nanodomains.

In **Fig. 8b**, as the fluence increases, more oxygen ions are introduced into the film, leading to a monotonic increase in the formation of Si–O and C–O bonds and a reduction of original bonds, such as Si–Si, Si–N, Si–C, and C–C. These results indicate that the bond network is progressively restructured with increasing plasma fluence, showing a linear relationship between plasma fluence and the extent of bond formation and dissociation. Since the rise in the amount of oxygen-related bonds involves a decrease in the density of the original bonds in the a-SiCN network, a higher plasma fluence could induce more significant chemical modifications to the surface. These composition- and fluence-dependent modifications in the bonding network were consistently observed across all examined a-SiCN compositions and plasma treatment conditions.

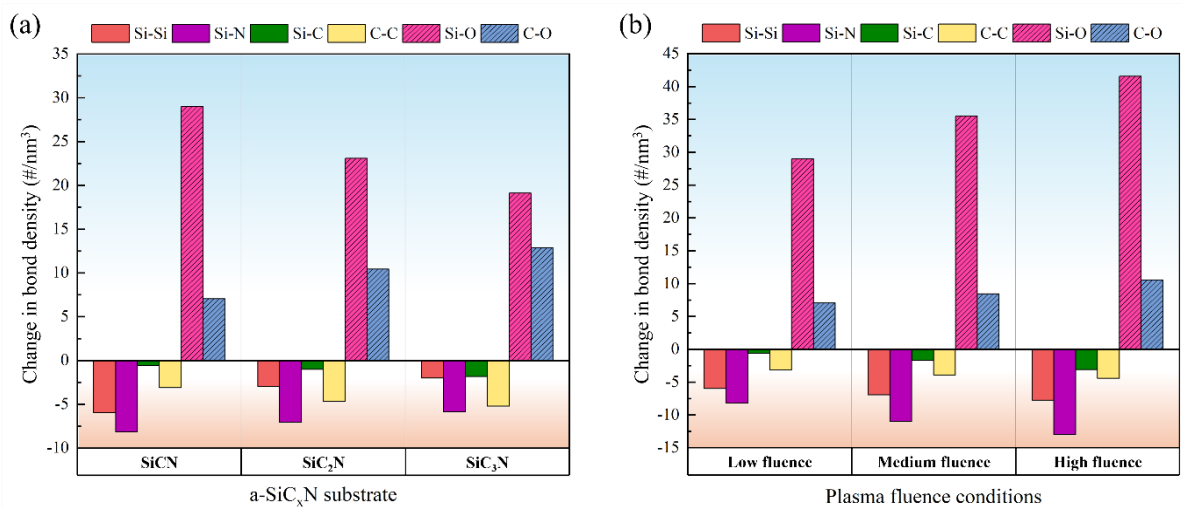

Fig. 8. Change in bond density of various bonds in a-SiCN films after $O_2$ plasma treatment, (a) depending on the SiCN composition at a fixed low plasma fluence condition, and (b) depending on plasma fluence for the a-SiCN model (Si:C:N = 1:1:1).

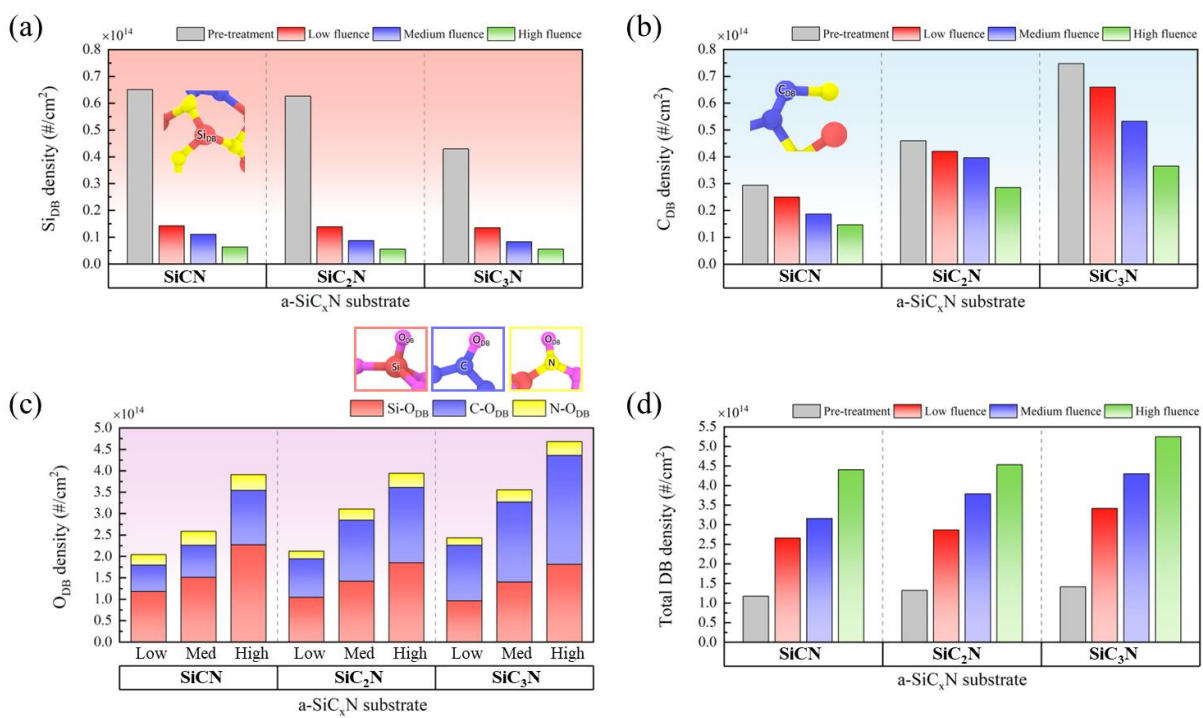

Fig. 9. Density of (a) $Si_{DB}$, (b) $C_{DB}$, (c) $O_{DB}$, and (d) total dangling bonds, including $Si_{DB}$, $C_{DB}$, $N_{DB}$, and $O_{DB}$, on a-$SiC_xN$ surfaces before and after plasma treatment. The insets are representative atomic structures of each dangling bond.

In **Fig. 9**, the dangling bond (DB) density was quantified within the effective depth of chemical modification ($z \geq 40\,\text{Å}$), and DBs were identified as undercoordinated atoms based on the ReaxFF bond order relative to the fully coordinated reference coordination number for each element.

Before plasma treatment, the DB density is dominated by $Si_{DB}$ (**Fig. 9a**) and $C_{DB}$ (**Fig. 9b**), whereas $N_{DB}$ is present only in relatively small amounts and changes negligibly with plasma treatment; accordingly, $N_{DB}$ is not depicted in **Fig. 9**. On the pre-treated surface, the densities of $Si_{DB}$ and $C_{DB}$ are approximately proportional to the Si and C content of the a-$SiC_xN$ film, as several surface atoms remain undercoordinated even after relaxation, resulting in the existence of $Si_{DB}$ and $C_{DB}$. Upon $O_2$ plasma exposure, the incoming ion species promote the formation of Si–O and C–O (**Fig. 8**) at these undercoordinated sites, converting a substantial fraction of $Si_{DB}$ and $C_{DB}$ into a fully coordinated state. Therefore, the $Si_{DB}$ and $C_{DB}$ densities are gradually reduced as the plasma fluence increases, as shown in **Fig. 9a** and **b**, respectively. Furthermore, the decrease in $Si_{DB}$ is significantly larger than that in $C_{DB}$ after plasma treatment, indicating that oxygen adsorption is favored at Si sites relative to C sites. This result is consistent with **Fig. 8**, which shows the increase in Si–O formation exceeds that of C–O across all cases. Such plasma-induced surface modification agrees well with XPS characterization of a-SiCN surfaces after $O_2$ plasma treatment by Kitagawa et al.,[14] who reported a pronounced increase in Si–O bonds with comparatively minor changes in C- and N-related bonds. It is worth noting that the decrease in $Si_{DB}$ and $C_{DB}$ could be interpreted as surface passivation, analogous to hydrogen plasma treatments;[50] however, **Fig. 9c** shows that $O_2$ plasma concurrently generates highly reactive $O_{DB}$ on the surface, and the $O_{DB}$ density is proportional to plasma fluence across all compositions. As shown in **Fig. 9d**, the plasma-treated surface exhibits a higher total DB density than the untreated surface; this net increase reflects surface activation by $O_2$ plasma treatment, where the growth of highly reactive $O_{DB}$ outweighs the concurrent reductions in $Si_{DB}$ and $C_{DB}$.

In **Fig. 9c**, $O_{DB}$ was classified by the element to which the oxygen is bonded, as Si–$O_{DB}$, C–$O_{DB}$, and N–$O_{DB}$. Although the N–$O_{DB}$ fraction is minor, the densities of Si–$O_{DB}$ and C–$O_{DB}$ depend on both the a-SiC$_x$N composition and plasma fluence: at a fixed composition, higher fluence increases all types of $O_{DB}$; at a fixed fluence, Si–$O_{DB}$ increases with the Si content, while C–$O_{DB}$ increases with the C content. The $O_{DB}$ density and the total DB density, depicted in **Fig. 9c** and **d**, respectively, rise as the C content in the a-SiC$_x$N substrate increases, mainly due to the growth of C–$O_{DB}$ density. Although DBs on the plasma-activated surface participate in forming interfacial linkages and consuming water generated at the interface, the Si–$O_{DB}$ density is especially important for bonding strength, because it acts as a potential source for forming Si–OH and subsequent Si–O–Si linkages.[14]

### 3.2.2 Surface hydroxylated single a-SiCN layer

In **Fig. 10a**, the density of silanol groups (Si–OH) formed after surface hydroxylation was quantified in the effective depth of chemical modification ($z \geq 40$ Å). **Fig. 10a** shows the Si–OH density increases with plasma fluence across all compositions, and a clear composition-dependence is also observed at each plasma fluence, SiCN > SiC$_2$N > SiC$_3$N. The a-SiCN surface plasma-treated at high fluence (**Fig. 10c**) has a greater number of Si–OH than at low fluence (**Fig. 10b**). At the same low fluence, a-SiCN (**Fig. 10b**) exhibits a denser Si–OH distribution than a-SiC$_3$N (**Fig. 10d**), reflecting the composition-dependence.

These fluence- and composition-dependent changes are related to the $O_2$ plasma-induced surface modification, characterized by the DB density. Since $Si_{DB}$ and $Si-O_{DB}$ react with water molecules to form Si–OH groups,[13,23,51] the measured Si–OH densities can be interpreted from the pre-hydroxylation populations of these precursors. As shown in **Fig. 9a**, the $Si_{DB}$ density depends on the SiCN composition and plasma fluence but remains substantially lower than the $Si-O_{DB}$ density (**Fig. 9c**). In **Fig. 10b–d**, Si–OH groups that contain an $H_2O$-derived oxygen correspond to hydroxylated $Si_{DB}$, and these are comparatively sparse relative to Si–OH originated from $Si-O_{DB}$. This observation is consistent with **Fig. 9a** and **c**, where the $Si-O_{DB}$ density is much higher than the $Si_{DB}$ density and implies that Si–OH population is primarily determined by the density of $Si-O_{DB}$ rather than $Si_{DB}$.

However, the Si–OH density is not directly proportional to the $Si-O_{DB}$ density (**Fig. 9c**). For instance, although the $Si-O_{DB}$ density in a-$SiC_3N$ at high fluence exceeds that in a-SiCN at low fluence, the Si–OH density is higher in the latter. These observations are attributed to different ways in which plasma treatment and surface hydroxylation deliver the reactants to the surface. During $O_2$ plasma treatment, energetic ions bombarded the surface and partially penetrated below the outermost layer, generating $O_{DB}$ within the effective depth of chemical modification within 1 nm-thickness. By contrast, surface hydroxylation proceeded through water–surface interactions at ambient temperature, without allowing penetration into the substrate. Accordingly, $Si-O_{DB}$ groups are distributed within the 1 nm-thickness near-surface region, whereas the Si–OH groups are found predominantly at the outermost surface, as shown in **Fig. 10b–d**. Furthermore, the distribution of C atoms near the surface hinders Si–OH formation by reducing the effective contact area between water molecules and Si-related DBs, an effect that becomes more pronounced as the C content increases, as depicted in **Fig. 10b–d**. Therefore, Si–OH formation reflects not only plasma surface activation but also the local Si/C arrangement at the surface, and the arrangement should be considered to obtain a more hydrophilic surface favorable for a robust bonding interface.

**Fig. 11** illustrates the morphological modification of each a-SiCN model induced by surface treatment, presenting the root mean square roughness ($R_q$) as a quantitative measure, along with the surface topography of the substrates. As shown in **Fig. 11a**, the pre-treatment surface roughness (0.088–0.093 nm) aligns well with values reported after chemical mechanical polishing (CMP), typically ranging from 0.09 to 0.1 nm.[13,16] Therefore, despite the absence of a CMP process in the atomistic modeling, the resulting flatness of the surface is comparable to that of post-CMP surfaces. This suggests that the atomistic models accurately reflect the effect of surface treatment on surface roughness.

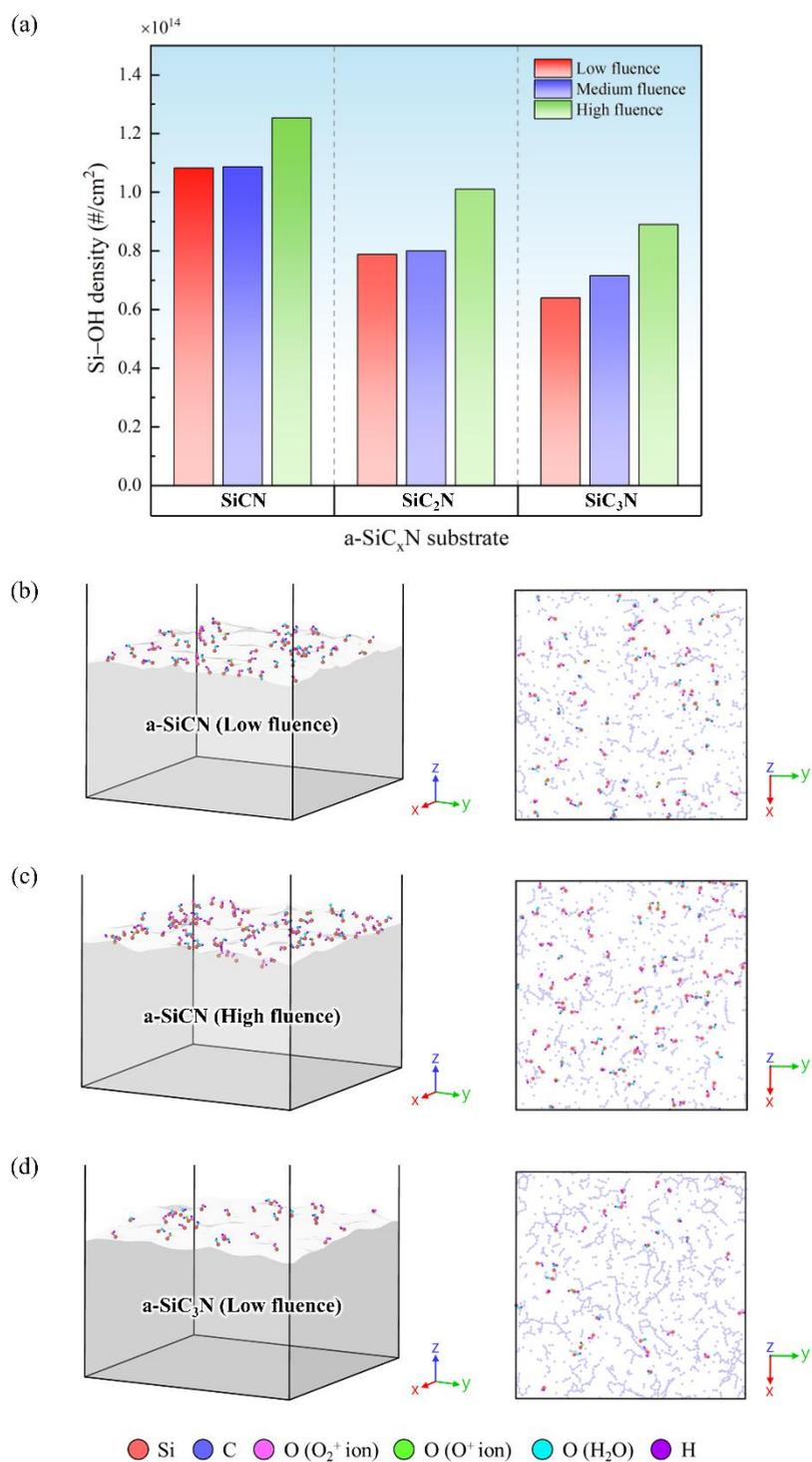

Fig. 10. (a) Areal density of Si–OH groups after the surface hydroxylation. Atomic configurations of the Si–OH groups for (b) a-SiCN, low fluence; (c) a-SiCN, high fluence; and (d) a-SiC$_3$N, low fluence. Left panels show perspective views with Si–OH groups, and right panels show top views with Si–OH groups overlaid with carbon atoms.

In **Fig. 11a** and **b**, the variation in surface roughness among different compositions is found to be negligible, within several $10^{-3}$ nm, indicating no significant composition-dependence under the same fluence condition. Contrarily, the surface roughness increases notably with increasing plasma fluence (**Fig. 11**), suggesting that a higher density of ion exposure induces more pronounced morphological modification. This fluence-dependent change in surface topography was also observed in experimental results, where enhanced surface roughening has been reported under higher plasma doses and longer periods of plasma exposure.[53] Although surface roughness was measured after the surface hydroxylation, the observed fluence-dependent variation reflects morphological modification that had already occurred during plasma treatment, caused by the accumulation of ion species on the surface.

Such morphological modification is known to adversely affect bonding performance by reducing the effective contact area and generating irregular bonding at the interface. Furthermore, an excessively rough surface is prone to generating interfacial voids and incomplete bonding for the dielectric bonding regime, which can reduce interfacial bonding strength and overall bonding reliability.[30] In this context, the increase in surface roughness observed at higher plasma fluence in our simulation is expected to act as a detrimental factor for bonding energy, potentially reducing interfacial adhesion.

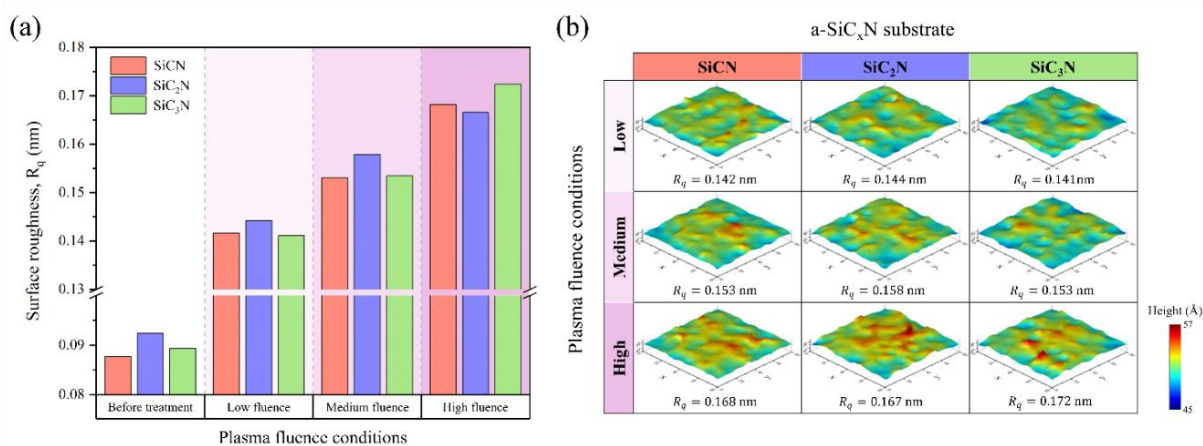

Fig. 11. (a) Surface roughness ($R_q$) of a-SiCN substrates before and after surface treatment with respect to SiCN composition and plasma fluence condition. (b) Surface topography of each a-SiCN model, presented with color scales representing surface height.

## 3.3 Bonding energy analysis

### 3.3.1 Bonding energy

**Fig. 12** presents the T–S responses and corresponding interfacial atomic configurations of three representative cells: Cells 2, 8, and 9, from the a-SiC$_2$N model under low fluence conditions. The exponential T–S curves exhibit nonlinear ductile debonding behavior, where traction increases with separation and then gradually decays to zero after complete interfacial separation. Atomic configurations (ⅰ)–(ⅲ) in **Fig. 12** illustrate the atomic trajectories during the debonding simulation, from the initial bonding interface to the complete separation. At the beginning of debonding, various interfacial bonds are observed at the bonding interface. These bonds resist tensile loading and progressively break during separation. In the decay region following the maximum traction, a few interfacial Si–O–Si linkages are observed in configurations (ⅱ), even after most interfacial bonds have broken. This observation indicates that Si–O–Si linkages are the primary structural components contributing to interfacial bonding strength in Si-based dielectric bonding schemes.

Notably, Cells 2, 8, and 9 exhibit saliently distinguishable T–S responses, particularly in terms of maximum traction and the point of complete separation. As shown in **Fig. 12b**, Cell 8 maintains traction over a wider separation range compared with Cell 2 (**Fig. 12a**), whereas Cell 9 (**Fig. 12c**) exhibits a narrower separation range with earlier traction decay. The atomic configuration (ⅱ) of Cell 8 reveals multiple Si–O–Si linkages connected in series, which act as robust interfacial bridges that delay complete separation to larger $\delta$. In contrast, the interfacial bonding network in Cell 9 is less prominent, resulting in a lower maximum traction and earlier complete separation. These inhomogeneous bonding characteristics arise from local variations in interfacial bonding configurations and the T–S responses capture these inhomogeneous bonding characteristics.

**Fig. 13** shows the exponential T–S curves and the resulting bonding energy for each cell. The mean bonding energy within each model, indicated by the dotted lines in **Fig. 13**, exhibits a clear dependence on both composition and plasma fluence: with respect to composition, a-SiCN > a-SiC$_2$N > a-SiC$_3$N; and with respect to plasma fluence, low fluence > medium fluence > high fluence. Since a similar dependence was observed in the structural characterization of plasma-treated surfaces, as discussed in **Section 3.2**, this observation implies a significant correlation between structural characteristics and bonding energy. However, this dependence is not uniformly maintained across all cells. For instance, Cell 8 in the a-SiC$_2$N model (**Fig. 13b**) shows a significantly higher bonding energy than any of the cells in the a-SiCN model (**Fig. 13a**). This deviation may result from the intrinsic non-uniformity of the surface in terms of local composition, bond density, and topography, suggesting the need to investigate the relationship between structural characterization and bonding energy at the cell level.

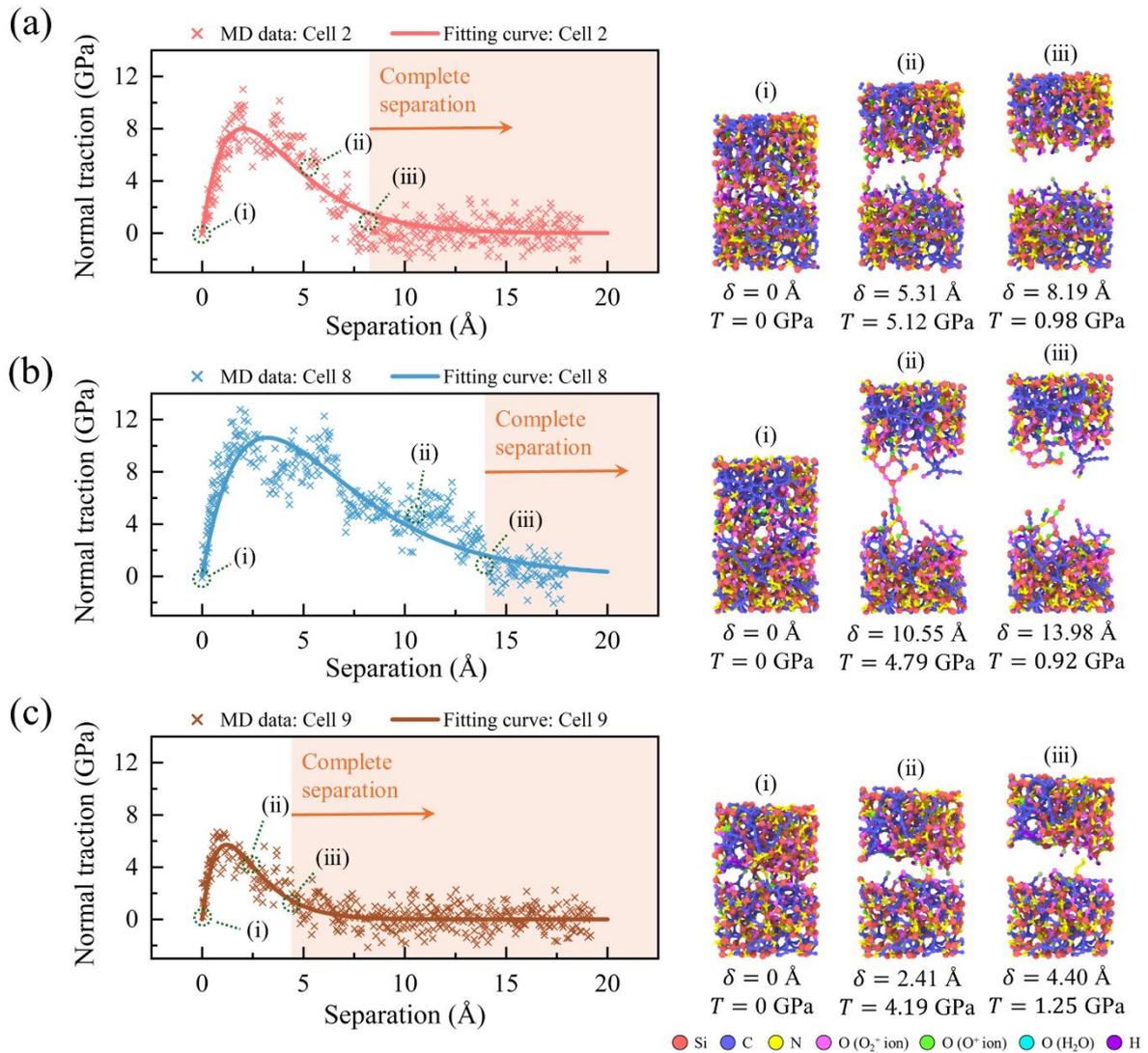

Fig. 12. Traction–separation data and exponential fitting curves for cells 2, 8, and 9 in the a-SiC$_2$N, low fluence model: (a) Cell 2; (b) Cell 8; (c) Cell 9. Atomic configurations (ⅰ)–(ⅲ) illustrate the initial state, the state with only a few interfacial bonds remaining, and immediately after complete separation, respectively.

The mean values of bonding energy obtained from the subdivided CZVE are 0.26–3.01 % higher than those obtained from the non-subdivided CZVE. This marginal discrepancy is attributed to fitting discrete atomistic T–S data to a continuous exponential law and is considered negligible for interface-average comparisons. Accordingly, subdividing the CZVE does not materially alter the overall bonding energy of the interface but enables the evaluation of locally resolved bonding energies that reflect the inhomogeneous bonding characteristics across the interface. The mean bonding energies range from 1.676 to 4.609 J/m$^2$, which agrees well with the experimentally reported values for SiCN-SiCN plasma-activated bonding (1.2–6.0 J/m$^2$).[14,23,30,51]

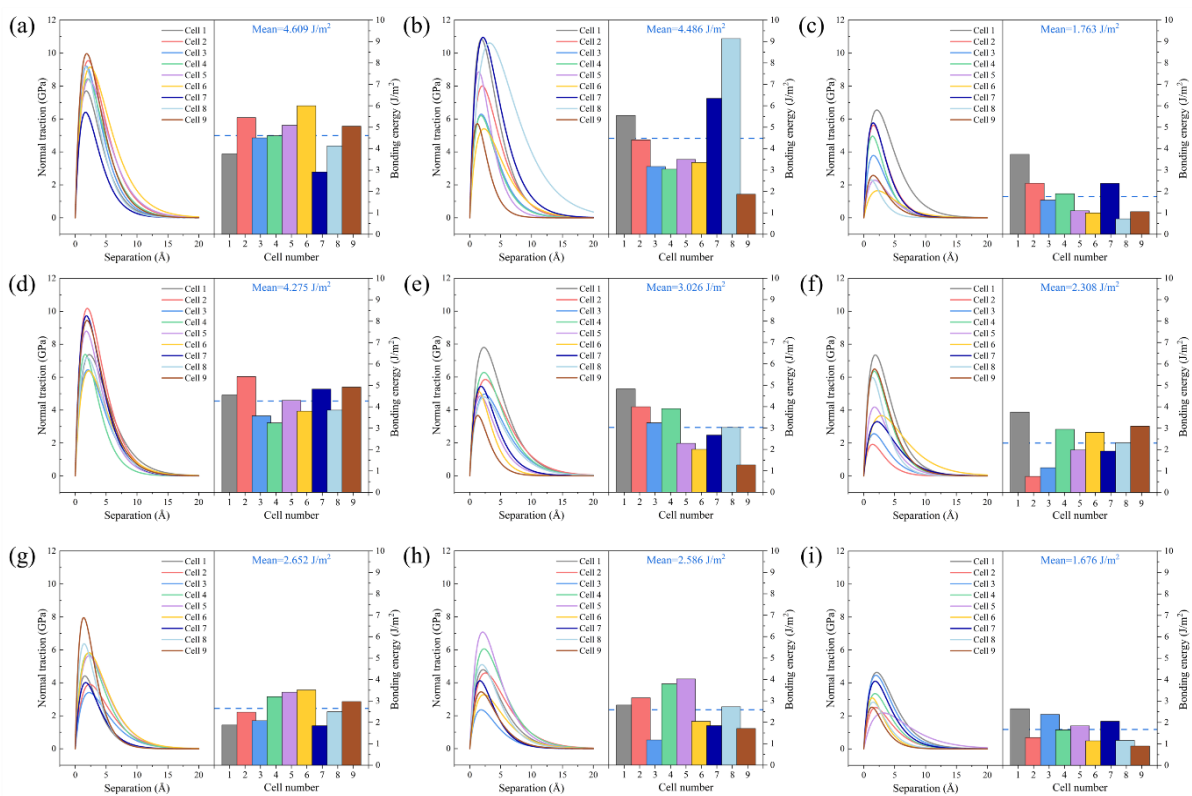

Fig. 13. Exponential traction–separation curves (left) of each cell and corresponding bonding energies (right): (a) a-SiCN, low fluence; (b) a-SiC$_2$N, low fluence; (c) a-SiC$_3$N, low fluence; (d) a-SiCN, medium fluence; (e) a-SiC$_2$N, medium fluence; (f) a-SiC$_3$N, medium fluence; (g) a-SiCN, high fluence; (h) a-SiC$_2$N, high fluence; (i) a-SiC$_3$N, high fluence. The dashed lines indicate the mean value for bonding energy.

*3.3.2 Correlation between bonding energy and structural characteristics*

In **Fig. 14**, the bonding energy obtained from each cell is plotted as a function of the interfacial Si–O–Si density. An overall positive correlation is observed between Si–O–Si density and bonding energy across different compositions and plasma fluence conditions, consistent with typical behavior in Si-based dielectric bonding interfaces. It indicates that higher interfacial Si–O–Si enhances interfacial connectivity and increases bonding energy. However, one data point corresponding to Cell 8 in the a-SiC$_2$N model exhibits a markedly higher bonding energy than other cells while having a relatively low Si–O–Si density. As observed in the atomic configuration (ii) of **Fig. 12b**, this distinctive case might be attributed to the presence of multiple Si–O–Si linkages connected in series, which lead to a higher bonding energy despite the lower Si–O–Si density. Such a localized bonding configuration was rarely observed in this study. Therefore, this data point was excluded from the construction of the 95 % confidence ellipses (shown as dashed curves in **Fig. 14a** and **b**), which visualize the data distributions according to SiCN composition and plasma fluence. The remaining data yield bonding energies of 0.72–6.35 J/m$^2$, comparable to the experimental range.

**Fig. 14a** shows the 95 % confidence ellipses corresponding to each composition are progressively shifted toward higher Si–O–Si density and higher bonding energy, following the order a-SiCN > a-SiC$_2$N > a-SiC$_3$N. The ellipses move upward and to the right with increasing Si content, while their principal-axis orientations remain similar, indicating that a higher Si content increases the mean level of Si–O–Si formation and bonding energy rather than changing the correlation slope. Such composition-dependence of bonding energy results from the surface Si–OH density observed in **Fig. 10a**. This relationship arises because Si–OH groups on the opposing surfaces undergo dehydration reactions at the bonding interface to form interfacial Si–O–Si linkages. Consequently, a-SiCN, which possesses a higher Si–OH density, yields overall higher bonding energy compared with a-SiC$_3$N. At comparable Si–O–Si densities, variation in bonding energy can be attributed to local morphological irregularities that alter the effective contact area between the surfaces.

**Fig. 14b** presents the influence of plasma fluence on the correlation between Si–O–Si density and bonding energy. The low fluence group generally exhibits higher bonding energy than the medium and high fluence groups, while maintaining a positive correlation in each case. Notably, the 95 % confidence ellipses for different fluence conditions share a comparable range of Si–O–Si density and partially overlap, indicating the fluence-dependent variation in bonding energy cannot be explained solely by the extent of Si–O–Si linkage formation associated with surface Si–OH density. We also found that the ellipse for the high fluence condition appears at a lower Si–O–Si density and bonding energy, which contradicts the structural characterization in **Fig. 10b** showing that surface Si–OH density increases with higher plasma fluence. The principal axes of ellipses become progressively steeper from high to low fluence, indicating a larger increase in bonding energy per unit increase in Si–O–Si density under lower fluence conditions. This difference in slope can be explained by the structural characterization results in **Fig. 11**, which show that higher plasma fluence increases surface roughness and morphological irregularity, thereby reducing the effective contact area and weakening interfacial connectivity. In contrast, the difference in surface roughness among compositions is negligible, which explains why the ellipses in **Fig. 14a** exhibit nearly identical orientations across all compositions. By subdividing the CZVE into cells, the local inhomogeneity of surface topography within each interface was effectively incorporated, enabling the fluence-dependent variations to be captured more distinctly.

It can therefore be concluded that **Fig. 14a** and **b** demonstrate the complementary roles of composition and plasma fluence: composition primarily shifts the mean position toward higher Si–O–Si density and higher bonding energy without significantly altering the correlation orientation, whereas plasma fluence changes the slope of the correlation, with lower fluence producing a steeper orientation and thus a greater per-unit increase in bonding energy for a given increase in Si–O–Si density. These findings highlight that the interfacial bonding properties of SiCN–SiCN bonded systems are determined by the combined effects of surface chemical and morphological modifications, providing atomistic insights useful for optimizing SiCN composition and plasma treatment parameters in plasma-activated bonding.

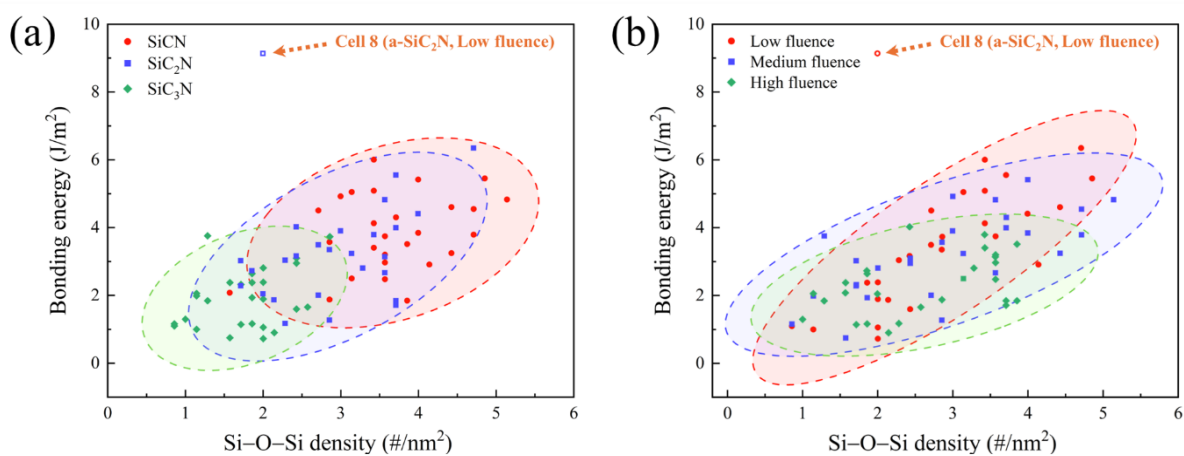

Fig. 14. Bonding energy of each cell as a function of the density of Si–O–Si linkages: (a) grouped by composition, (b) grouped by plasma fluence. The dashed curves denote the 95 % confidence ellipses, and the bonding energy of Cell 8 in the a-SiC$_2$N model under low fluence conditions was not used to construct these ellipses.

## 4. Conclusion

In this study, we investigated the correlation between structural characteristics and the bonding energy in the SiCN–SiCN plasma-activated bonding scheme from an atomistic perspective. We considered various SiCN compositions and plasma fluence conditions to examine their effects on surface modification. The entire bonding sequence was performed by using reactive molecular dynamics simulations, which capture bond formation and dissociation dynamically through bond order evolution throughout the process. After the bonding sequence, the exponential traction–separation (T–S) response was obtained during the debonding simulation within subdivided cohesive zone volume elements (CZVE), from which the bonding energy was evaluated.

First, we observed that the a-SiCN surfaces exhibit chemical and morphological modifications after the surface treatment. Chemically, the number of surface dangling bonds increased with plasma fluence and depended on composition. Morphologically, surface roughness increased with plasma fluence, reducing the contact area at the bonding interface and interfacial connectivity. Both types of surface modification show clear dependence on composition and plasma fluence, which exert opposing effects on the interfacial Si–O–Si density and consequently affect the bonding energy that is strongly correlated with the Si–O–Si density. This finding suggests that both SiCN composition and plasma treatment parameters should be simultaneously considered to enhance interfacial bonding strength.

The present study establishes an atomic-level material–process–property relationship for SiCN–SiCN plasma-activated bonding. We believe that these outcomes provide valuable practical guidance in enhancing the bonding strength for plasma-activated bonding of SiCN films, which are essential for developing a novel hybrid bonding strategy. This framework is readily transferable to related dielectric–dielectric bonding scenarios and offers a quantitative basis for process design.

**Acknowledgments**

This research was supported by the National R&D Program through the National Research Foundation of Korea (NRF), funded by the Ministry of Science and ICT (No. RS-2025-23525252), and by the InnoCORE program of the Ministry of Science and ICT (No. N10250154). This work was also supported by the Korea Institute of Energy Technology Evaluation and Planning (KETEP) and the Ministry of Trade, Industry & Energy (MOTIE) of the Republic of Korea (No. RS-2023-00240918).

**References**

(1) Lau, J. H. State of the Art of Cu-Cu Hybrid Bonding. *IEEE Trans. Components, Packag. Manuf. Technol.* **2024**, *14* (3), 376–396. https://doi.org/10.1109/TCPMT.2024.3367985.

(2) Kang, Q.; Li, G.; Li, Z.; Tian, Y.; Wang, C. Surface Co-Hydrophilization via Ammonia Inorganic Strategy for Low-Temperature Cu/SiO2 Hybrid Bonding. *J. Mater. Sci. Technol.* **2023**, *149*, 161–166. https://doi.org/10.1016/j.jmst.2022.12.012.

(3) Lu, M. C. Hybrid Bonding for Ultra-High-Density Interconnect. *J. Electron. Packag.* **2024**, *146* (3), 31–37. https://doi.org/10.1115/1.4064750.

(4) Lau, J. H. Recent Advances and Trends in Cu-Cu Hybrid Bonding. *IEEE Trans. Components, Packag. Manuf. Technol.* **2023**, *13* (3), 399–425. https://doi.org/10.1109/TCPMT.2023.3265529.

(5) Sakuma, K.; Yu, R.; Polomoff, N.; Kumar, A.; Raghavan, S.; Belyansky, M.; Jain, A.; Bonam, R.; Sulehria, Y.; Hsu, H.; Jayanand, K.; Jagannathan, H. D2W and W2W Hybrid Bonding System with below 2.5 Micron Pitch for 3D Chiplet AI Applications. *Tech. Dig. - Int. Electron Devices Meet. IEDM* **2024**, 1–4. https://doi.org/10.1109/IEDM50854.2024.10873522.

(6) Hu, H. W.; Chen, K. N. Development of Low Temperature Cu[Sbnd]Cu Bonding and Hybrid Bonding for Three-Dimensional Integrated Circuits (3D IC). *Microelectron. Reliab.* **2021**, *127* (October), 114412. https://doi.org/10.1016/j.microrel.2021.114412.

(7) Onishi, K.; Kitagawa, H.; Teranishi, S.; Uedono, A.; Inoue, F. Temporary Direct

Bonding by Low Temperature Deposited SiO2 for Chiplet Applications. *ACS Appl. Electron. Mater.* **2024**, *6* (4), 2449–2456. https://doi.org/10.1021/acsaelm.4c00114.

(8) Tabata, T.; Sanchez, L.; Larrey, V.; Fournel, F.; Moriceau, H. SiO2-SiO2 Die-to-Wafer Direct Bonding Interface Weakening. *Microelectron. Reliab.* **2020**, *107* (February), 2–6. https://doi.org/10.1016/j.microrel.2020.113589.

(9) He, R.; Fujino, M.; Yamauchi, A.; Suga, T. Combined Surface Activated Bonding Technique for Hydrophilic SiO 2 -SiO 2 and Cu-Cu Bonding . *ECS Meet. Abstr.* **2016**, *MA2016-02* (32), 2082–2082. https://doi.org/10.1149/ma2016-02/32/2082.

(10) Zhang, X. X.; Raskin, J. P. Low-Temperature Wafer Bonding: A Study of Void Formation and Influence on Bonding Strength. *J. Microelectromechanical Syst.* **2005**, *14* (2), 368–382. https://doi.org/10.1109/JMEMS.2004.839027.

(11) Fournel, F.; Martin-Cocher, C.; Radisson, D.; Larrey, V.; Beche, E.; Morales, C.; Delean, P. A.; Rieutord, F.; Moriceau, H. Water Stress Corrosion in Bonded Structures. *ECS J. Solid State Sci. Technol.* **2015**, *4* (5), P124–P130. https://doi.org/10.1149/2.0031505jss.

(12) Yang, Y.; Brun, X. F.; Weber, M. H.; Flores, M. Quantification of Interfacial Voids Using Positron Annihilation Spectroscopy for Mechanism Study on SiCN Bonding and SiN Bonding. *ECS J. Solid State Sci. Technol.* **2024**, *13* (11), 113002. https://doi.org/10.1149/2162-8777/ad8c82.

(13) Inoue, F.; Peng, L.; Iacovo, S.; Phommahaxay, A.; Verdonck, P.; Meersschaut, J.; Dara, P.; Sleeckx, E.; Miller, A.; Beyer, G.; Beyne, E. Influence of Composition of SiCN as Interfacial Layer on Plasma Activated Direct Bonding. *ECS J. Solid State Sci. Technol.* **2019**, *8* (6), P346–P350. https://doi.org/10.1149/2.0241906jss.

(14) Kitagawa, H.; Sato, R.; Ebiko, S.; Nagata, A.; Ahn, C.; Kim, Y.; Kang, J.; Uedono, A.; Inoue, F. Material-Mechanistic Interplay in SiCN Wafer Bonding for 3D Integration. *ACS Omega* **2025**, *10* (25), 27575–27584. https://doi.org/10.1021/acsomega.5c03628.

(15) Nakayama, K.; Hayama, K.; Tanaka, F. L.; La, M. T. N.; Inoue, F. Minimizing Recess of Cu Pad on Hybrid Bonding with SiCN via Non-Selective Chemical Mechanical Polishing and Post-Cleaning Steps. *ECS J. Solid State Sci. Technol.* **2024**, *13* (7),

074009. https://doi.org/10.1149/2162-8777/ad5fb7.

(16) Nagano, F.; Inoue, F.; Phommahaxay, A.; Peng, L.; Chancerel, F.; Naser, H.; Beyer, G.; Uedono, A.; Beyne, E.; De Gendt, S.; Iacovo, S. Origin of Voids at the SiO 2 /SiO 2 and SiCN/SiCN Bonding Interface Using Positron Annihilation Spectroscopy and Electron Spin Resonance . *ECS J. Solid State Sci. Technol.* **2023**, *12* (3), 033002. https://doi.org/10.1149/2162-8777/acbe18.

(17) Brun, X. F.; Hasan, M. M.; Yang, Y.; Carazzetti, P.; Drechsel, C.; Strolz, E. Characterization of 300 Mm Low Temperature SiCN PVD Films for Hybrid Bonding Application. *Proc. - Electron. Components Technol. Conf.* **2023**, *2023-May*, 548–555. https://doi.org/10.1109/ECTC51909.2023.00098.

(18) Sato, R.; Nagata, A.; Kitagawa, H.; Kondo, Y.; Saito, K.; Park, J.; Ahn, C.; Kim, Y. S.; Kang, J.; Inoue, F. Bond Wave Analysis of SiCN for Fine Pitch Hybrid Bonding. *Proc. - Electron. Components Technol. Conf.* **2025**, 2046–2053. https://doi.org/10.1109/ECTC51687.2025.00350.

(19) Inoue, F.; Nagata, A.; Fuse, J.; Ebiko, S.; Sato, R.; Saito, K.; Kondo, Y.; Kawauchi, T.; Park, J.; Ahn, C.; Kim, M.; Kang, J. Low Temperature Wafer Level Hybrid Bonding Enabled by Advanced SiCN and Surface Activation. *Proc. - Electron. Components Technol. Conf.* **2024**, 69–75. https://doi.org/10.1109/ECTC51529.2024.00020.

(20) Van Duin, A. C. T.; Dasgupta, S.; Lorant, F.; Goddard, W. A. ReaxFF: A Reactive Force Field for Hydrocarbons. *J. Phys. Chem. A* **2001**, *105* (41), 9396–9409. https://doi.org/10.1021/jp004368u.

(21) Senftle, T. P.; Hong, S.; Islam, M. M.; Kylasa, S. B.; Zheng, Y.; Shin, Y. K.; Junkermeier, C.; Engel-Herbert, R.; Janik, M. J.; Aktulga, H. M.; Verstraelen, T.; Grama, A.; Van Duin, A. C. T. The ReaxFF Reactive Force-Field: Development, Applications and Future Directions. *npj Comput. Mater.* **2016**, *2* (September 2015). https://doi.org/10.1038/npjcompumats.2015.11.

(22) Hahn, S. H.; Kim, W.; Shin, D.; Lee, Y.; Choi, W.; Moon, B.; Lim, K.; Rhee, M. Molecular Dynamics Study on Plasma-Surface Interactions of SiCN Dielectrics for Wafer-to-Wafer Hybrid Bonding Process. *Proc. 24th Electron. Packag. Technol. Conf. EPTC 2022* **2022**, 527–533. https://doi.org/10.1109/EPTC56328.2022.10013205.

(23) Kim, H.; Tuchman, A.; Tsai, Y. H.; Hisamatsu, T.; Son, I.; Arkalgud, S. Atomistic Simulation Investigation of Various Plasma Surface Activations in SiCN Dielectric Bonding. *Proc. - Electron. Components Technol. Conf.* **2024**, 1633–1638. https://doi.org/10.1109/ECTC51529.2024.00269.

(24) Liao, N.; Xue, W.; Zhang, M. Effect of Carbon Content on Structural and Mechanical Properties of SiCN by Atomistic Simulations. *J. Eur. Ceram. Soc.* **2012**, *32* (6), 1275–1281. https://doi.org/10.1016/j.jeurceramsoc.2011.11.022.

(25) Tersoff, J. Modeling Solid-State Chemistry: Interatomic Potentials for Multicomponent Systems. *Phys. Rev. B* **1989**, *39* (8), 5566–5568. https://doi.org/10.1103/PhysRevB.39.5566.

(26) de Brito Mota, F.; Justo, J.; Fazzio, A. Structural Properties of Amorphous Silicon Nitride. *Phys. Rev. B - Condens. Matter Mater. Phys.* **1998**, *58* (13), 8323–8328. https://doi.org/10.1103/PhysRevB.58.8323.

(27) Joshi, K. L.; Psofogiannakis, G.; Van Duin, A. C. T.; Raman, S. Reactive Molecular Simulations of Protonation of Water Clusters and Depletion of Acidity in H-ZSM-5 Zeolite. *Phys. Chem. Chem. Phys.* **2014**, *16* (34), 18433–18441. https://doi.org/10.1039/c4cp02612h.

(28) Sun, H.; Hu, Y.; Bai, L.; Xu, J. ReaxFF Molecular Dynamics Simulation of Single-Crystalline Silicon Plasma Polishing and Subsurface Damage Removal. *Comput. Mater. Sci.* **2024**, *233* (June 2023), 112685. https://doi.org/10.1016/j.commatsci.2023.112685.

(29) Yeon, J.; Chowdhury, S. C.; Gillespie, J. W. Hydroxylation and Water-Surface Interaction for S-Glass and Silica Glass Using ReaxFF Based Molecular Dynamics Simulations. *Appl. Surf. Sci.* **2023**, *608* (May 2022), 155078. https://doi.org/10.1016/j.apsusc.2022.155078.

(30) Chidambaram, V.; Lianto, P.; Wang, X.; See, G.; Wiswell, N.; Kawano, M. Dielectric Materials Characterization for Hybrid Bonding. *Proc. - Electron. Components Technol. Conf.* **2021**, *2021-June*, 426–431. https://doi.org/10.1109/ECTC32696.2021.00078.

(31) Ajayi, T.; Gatimu, A.; Brun, X. F.; Woods, C.; Song, K.; Bhat, A.; Galande, C.; Le, P.;

Karim, Z. Impact of Thermal Annealing and Other Process Parameters on Hybrid Bonding Performance for 3D Advanced Assembly Technology. *Proc. - Electron. Components Technol. Conf.* **2023**, *2023-May*, 1426–1432. https://doi.org/10.1109/ECTC51909.2023.00243.

(32) Thompson, P. B.; Johnson, R.; Nadimpalli, S. P. V. Effect of Temperature on the Fracture Behavior of Cu/SAC305/Cu Solder Joints. *Eng. Fract. Mech.* **2018**, *199* (June), 730–738. https://doi.org/10.1016/j.engfracmech.2018.07.004.

(33) Kim, M. K.; Park, S.; Jang, A.; Lee, H.; Baek, S.; Lee, C. S.; Kim, I.; Park, J.; Jee, Y.; Kang, U. B.; Kim, D. W. Characterization of Die-to-Wafer Hybrid Bonding Using Heterogeneous Dielectrics. *Proc. - Electron. Components Technol. Conf.* **2022**, *2022-May*, 335–339. https://doi.org/10.1109/ECTC51906.2022.00062.

(34) Wang, R.; Han, J.; Mao, J.; Hu, D.; Liu, X.; Guo, X. A Molecular Dynamics Based Cohesive Zone Model for Interface Failure under Monotonic Tension of 3D Four Direction SiCf/SiC Composites. *Compos. Struct.* **2021**, *274* (June), 114397. https://doi.org/10.1016/j.compstruct.2021.114397.

(35) Lee, S.; Park, J.; Yang, J.; Lu, W. Molecular Dynamics Simulations of the Traction-Separation Response at the Interface between PVDF Binder and Graphite in the Electrode of Li-Ion Batteries. *J. Electrochem. Soc.* **2014**, *161* (9), A1218–A1223. https://doi.org/10.1149/2.0051409jes.

(36) Needleman, A. An Analysis of Decohesion along an Imperfect Interface. *Int. J. Fract.* **1990**, *42* (1), 21–40. https://doi.org/10.1007/BF00018611.

(37) Son, S.; Min, J.; Jung, E.; Kim, H.; Kim, T.; Jeon, H.; Kim, J.; Kim, S.; Moon, K.; Na, H.; Hwang, K.; Yeom, G. Y. Characteristics of Plasma-Activated Dielectric Film Surfaces for Direct Wafer Bonding. *Proc. - Electron. Components Technol. Conf.* **2020**, *2020-June*, 2025–2032. https://doi.org/10.1109/ECTC32862.2020.00315.

(38) Yuan, X.; Cormack, A. N. Si-O-Si Bond Angle and Torsion Angle Distribution in Vitreous Silica and Sodium Silicate Glasses. *J. Non. Cryst. Solids* **2003**, *319* (1–2), 31–43. https://doi.org/10.1016/S0022-3093(02)01960-9.

(39) Schempp, S.; Dürr, J.; Lamparter, P.; Bill, J.; Aldinger, F. Study of the Atomic Structure and Phase Separation in Amorphous Si-C-N Ceramics by X-Ray and Neutron Diffraction. *Zeitschrift fur Naturforsch. - Sect. A J. Phys. Sci.* **1998**, *53* (3–4), 127–133. https://doi.org/10.1515/zna-1998-3-405.

(40) Ivashchenko, V. I.; Kozak, A. O.; Porada, O. K.; Ivashchenko, L. A.; Sinelnichenko, O. K.; Lytvyn, O. S.; Tomila, T. V.; Malakhov, V. J. Characterization of SiCN Thin Films: Experimental and Theoretical Investigations. *Thin Solid Films* **2014**, *569* (C), 57–63. https://doi.org/10.1016/j.tsf.2014.08.027.

(41) Resta, N.; Kohler, C.; Trebin, H. R. Molecular Dynamics Simulations of Amorphous Si-C-N Ceramics: Composition Dependence of the Atomic Structure. *J. Am. Ceram. Soc.* **2003**, *86* (8), 1409–1414. https://doi.org/10.1111/j.1151-2916.2003.tb03484.x.

(42) Matsunaga, K.; Iwamoto, Y.; Fisher, C. A. J.; Matsubara, H. Molecular Dynamics Study of Atomic Structures in Amorphous Si-C-N Ceramics. *J. Ceram. Soc. Japan* **1999**, *107* (11), 1025–1031. https://doi.org/10.2109/jcersj.107.1025.

(43) Lehmann, G.; Hess, P.; Wu, J. J.; Wu, C. T.; Wong, T. S.; Chen, K. H.; Chen, L. C.; Lee, H. Y.; Amkreutz, M.; Frauenheim, T. Structure and Elastic Properties of Amorphous Silicon Carbon Nitride Films. *Phys. Rev. B* **2001**, *64* (16), 1653051–16530510. https://doi.org/10.1103/PhysRevB.64.165305.

(44) Chattopadhyay, S.; Chen, L. C.; Wu, C. T.; Chen, K. H.; Wu, J. S.; Chen, Y. F.; Lehmann, G.; Hess, P. Thermal Diffusivity in Amorphous Silicon Carbon Nitride Thin Films by the Traveling Wave Technique. *Appl. Phys. Lett.* **2001**, *79* (3), 332–334. https://doi.org/10.1063/1.1386619.

(45) Abdul Rahman, M. A.; Chiu, W. S.; Haw, C. Y.; Badaruddin, R.; Tehrani, F. S.; Rusop, M.; Khiew, P.; Rahman, S. A. Multi-Phase Structured Hydrogenated Amorphous Silicon Carbon Nitride Thin Films Grown by Plasma Enhanced Chemical Vapour Deposition. *J. Alloys Compd.* **2017**, *721*, 70–79. https://doi.org/10.1016/j.jallcom.2017.05.289.

(46) Badaruddin, M. R.; Muhamad, M. R.; Rahman, S. A. Multi-Phase Structured Silicon Carbon Nitride Thin Films Prepared by Hot-Wire Chemical Vapour Deposition. *Thin Solid Films* **2011**, *519* (15), 5082–5085. https://doi.org/10.1016/j.tsf.2011.01.133.

(47) A. Bendeddouche, R. Berjoan, E. Bêche, T. Merle-Mejean, S. Schamm, V. Serin, G. Taillades, A. Pradel, R. H. Structural Characterization of Amorphous SiCxNy

Chemical Vapor Deposited Coatings. *J. Appl. Phys.* **1997**, *81* (February), 6147–6154.

(48) Reinold, L. M.; Graczyk-Zajac, M.; Gao, Y.; Mera, G.; Riedel, R. Carbon-Rich SiCN Ceramics as High Capacity/High Stability Anode Material for Lithium-Ion Batteries. *J. Power Sources* **2013**, *236*, 224–229. https://doi.org/10.1016/j.jpowsour.2013.02.046.

(49) Jiang, T.; Wang, Y.; Wang, Y.; Orlovskaya, N.; An, L. Quantitative Raman Analysis of Free Carbon in Polymer-Derived Ceramics. *J. Am. Ceram. Soc.* **2009**, *92* (10), 2455–2458. https://doi.org/10.1111/j.1551-2916.2009.03233.x.

(50) Kim, B.; Kim, M.; Yoo, S.; Nam, S. K. Atomistic Insights on Hydrogen Plasma Treatment for Stabilizing High-k/Si Interface. *Appl. Surf. Sci.* **2022**, *593* (March 2022), 153297. https://doi.org/10.1016/j.apsusc.2022.153297.

(51) Lee, R. J.; He, P. S.; Tran, D. P.; Chiu, W. L.; Chang, H. H.; Lee, C. C.; Chen, C. Surface Modification of Nanotwinned Copper and SiCN Using N2 and Ar Plasma Activation. *Appl. Surf. Sci.* **2025**, *684* (November 2024), 161832. https://doi.org/10.1016/j.apsusc.2024.161832.

(52) Nagano, F.; Iacovo, S.; Phommahaxay, A.; Inoue, F.; Sleeckx, E.; Beyer, G.; Beyne, E.; De. Gendt, S. Film Characterization of Low-Temperature Silicon Carbon Nitride for Direct Bonding Applications. *ECS J. Solid State Sci. Technol.* **2020**, *9* (12), 123011. https://doi.org/10.1149/2162-8777/abd260.

(53) Alam, A. U.; Howlader, M. M. R.; Deen, M. J. The Effects of Oxygen Plasma and Humidity on Surface Roughness, Water Contact Angle and Hardness of Silicon, Silicon Dioxide and Glass. *J. Micromechanics Microengineering* **2014**, *24* (3). https://doi.org/10.1088/0960-1317/24/3/035010.